\documentclass[sigconf]{acmart}

\AtBeginDocument{%
  }

\copyrightyear{2025}
\acmYear{2025}
\setcopyright{acmlicensed}\acmConference[CHI '25]{CHI Conference on Human Factors in Computing Systems}{April 26-May 1, 2025}{Yokohama, Japan}
\acmBooktitle{CHI Conference on Human Factors in Computing Systems (CHI '25), April 26-May 1, 2025, Yokohama, Japan}
\acmDOI{10.1145/3706598.3713102}
\acmISBN{979-8-4007-1394-1/25/04}

\acmSubmissionID{7318}

\begin{document}

% See https://capitalizemytitle.com/ for capitalization
% \title{Navigating Complexity: Exploring the Speed-Accuracy Tradeoff in Curved Path Steering}
\title{Curves Ahead: Enhancing the Steering Law for Complex Curved Trajectories}

\author{Jennie J.Y. Chen}
\affiliation{%
  \institution{University of British Columbia}
  \city{Vancouver}
  \country{Canada}
  }
\email{jchn@ece.ubc.ca}

\author{Sidney S. Fels}
\affiliation{%
  \institution{University of British Columbia}
  \city{Vancouver}
  \country{Canada}
  }
\email{ssfels@ece.ubc.ca}

\renewcommand{\shortauthors}{Chen et al.}

\begin{abstract}
The Steering Law has long been a fundamental model in predicting movement time for tasks involving navigating through constrained paths,
such as in selecting sub-menu options,
particularly for straight and circular arc trajectories.
However, this does not reflect the complexities of real-world tasks where curvatures can vary arbitrarily, limiting its applications.
This study aims to address this gap by introducing the total curvature parameter \(K\) into the equation to account for the overall curviness characteristic of a path.
To validate this extension, we conducted a mouse-steering experiment on fixed-width paths with varying lengths and curviness levels.
Our results demonstrate that the introduction of \(K\) significantly improves model fitness for movement time prediction over traditional models.
These findings advance our understanding of movement in complex environments and support potential applications in fields like speech motor control and virtual navigation.

\end{abstract}





\maketitle

\noindent
\textbf{\small © {Owner/Author | ACM} 2025. This is the author's version of the work. It is posted here for your personal use. Not for redistribution. The definitive Version of Record was published in CHI ’25, https://doi.org/10.1145/3706598.3713102}
\vspace{10pt}

\section{Introduction}
\begin{figure}[tb]
  \centering
  \includegraphics[width=0.8\linewidth]{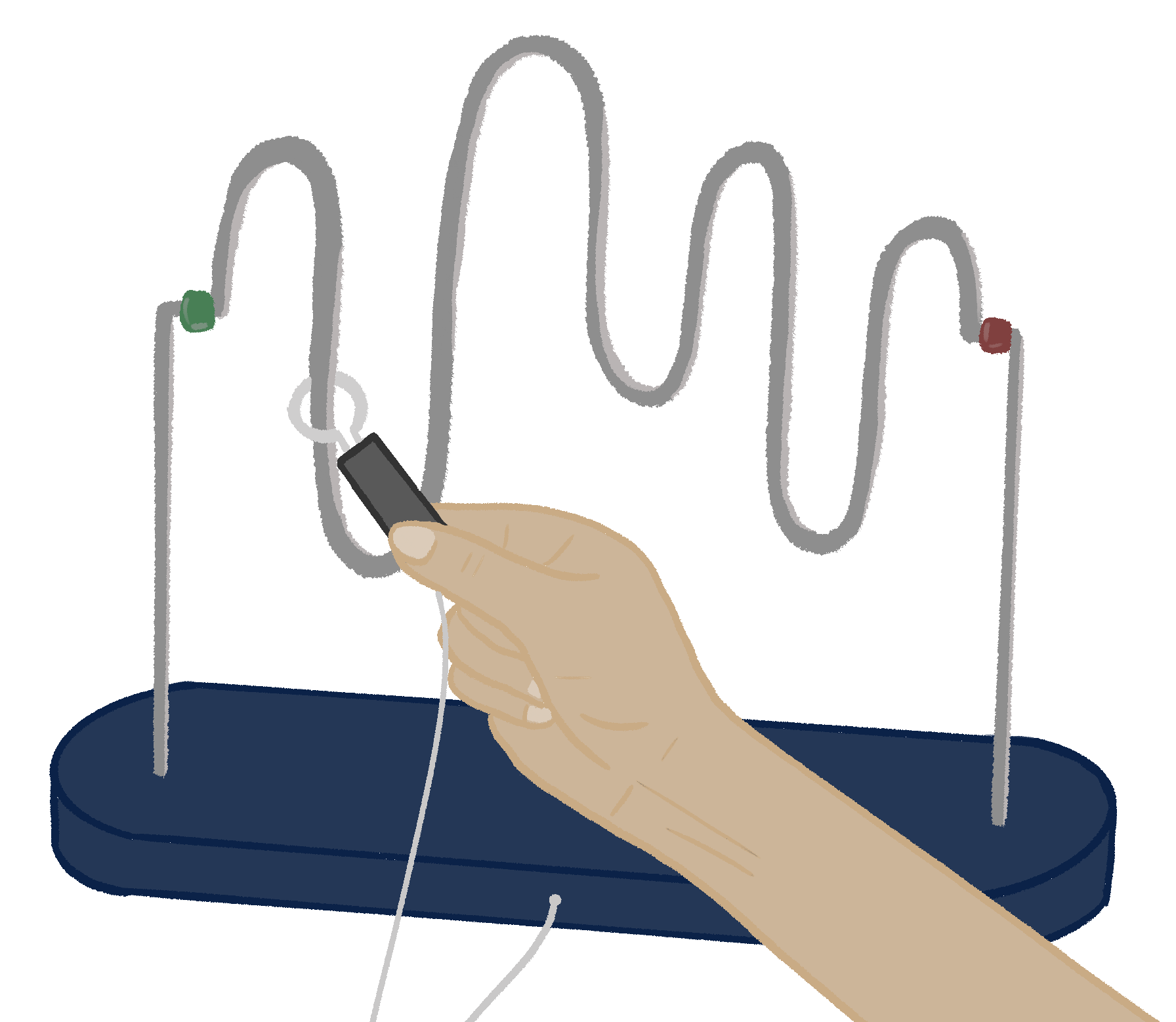}
  \caption{A children's game akin to a steering task. The goal is to move a metal loop from one end of a wire to the other without the two components touching each other.}
  \Description{This figure illustrates a children’s game that models a basic steering task. The image shows a human hand holding a metal loop, maneuvering it along a curvy wire path mounted on a base. The task requires guiding the loop from one end of the wire to the other without contact between the elements, simulating the challenges of navigating through complex paths in digital environments.}
  \label{fig_wire_game}
\end{figure}

Consider a children's game where the objective is to guide a metal loop along a curvy wire without touching it as shown in Figure \ref{fig_wire_game}.
The difficulty of this task is directly influenced by the length of the path, the size of the loop, and the bends in the wire, but how much more difficult does it get if we straightened the wire or added an extra bend?
The game, while simple, illustrates a broader problem in human performance modeling.

In Human-Computer Interaction (HCI) research, two models are commonly used in movement time prediction and task difficulty evaluation.
The most prominent one, Fitts' Law, describes the movement time taken to perform reciprocal pointing tasks based on the width of the targets and the distance between them \cite{fitts_information_1954}.
One formulation of Fitts' Law is derived from Shannon's Theorem 17 in information theory \cite{shannon_mathematical_1948}.
In this analogy, information is transmitted through the human channel, and movements are assigned an index of difficulty (\(ID\), given in bits) based on the task's parameters \cite{mackenzie_fitts_1992}.
The amplitude of the movement needed to perform a pointing task is analogous to signal power, and the width constraint of the pointing target is analogous to noise power, limiting the movement time - the information capacity of the human motor system.

The second human performance model, the Steering Law, proposed by Accot and Zhai, extends Fitts’ Law to model trajectory-based tasks where users must steer through a constrained tunnel \cite{accot_beyond_1997}.
Once again, the tunnel's length and width constraints are used to determine the index of difficulty of the task.

\begin{figure}[htb]
  \centering
  \includegraphics[width=\linewidth]{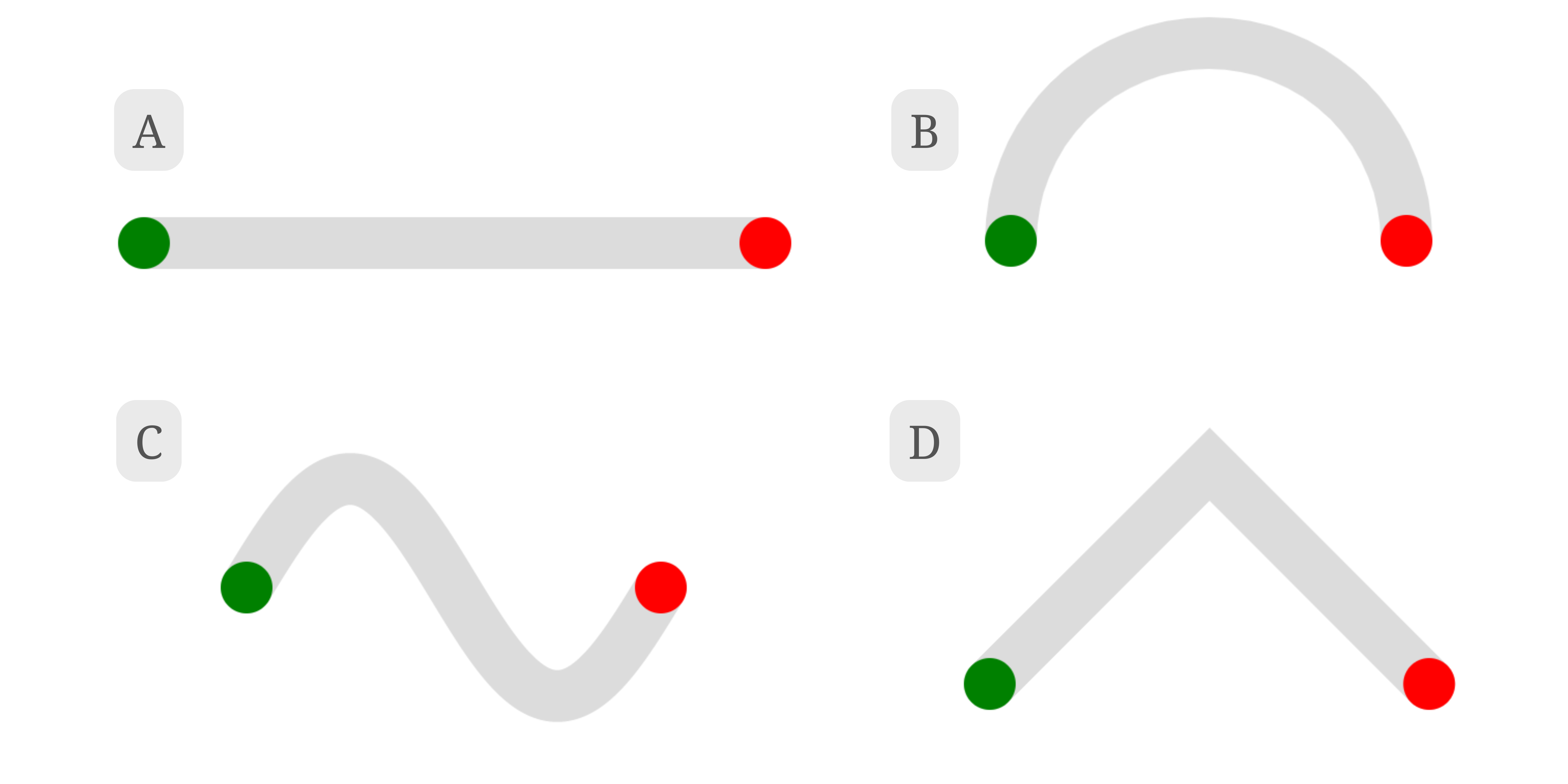}
  \caption{Different types of paths with the same width and length. A) Straight; B) Circular arc; C) Sinusoidal; D) Path with corner.}
  \label{fig_tunnels}
  \Description{This figure shows four different types of paths, each with identical width and length but varying in shape. Panel A shows a straight horizontal path, Panel B shows a semicircle arc path, Panel C shows a sinusoidal path with one complete period, and Panel D shows a sharp angled path with a 90 degree corner. All paths are marked with a green circle at the starting point on the left and a red circle at the endpoint on the right.}
\end{figure} 

In HCI research, Fitts' Law and Steering Law are often applied to user interface design such as navigating through drop-down menus and designing button interactions, evaluating the performance of input devices \cite{soukoreff_towards_2004, accot_performance_1999, zabramski_easy_2013} and comparing the performance of user groups \cite{welford_speed_1969, zhou_assessing_2011}.
There are also proposed models that combine the two fundamental laws for compound tasks that involve targeted steering actions \cite{dennerlein_force_2000, kulikov_targeted_2006, hong_human_2007}.
However, as interaction tasks become more complex, the limitations of these fundamental models become apparent.
For instance, all of the paths illustrated in Figure \ref{fig_tunnels} would be assigned the same \(ID\) by the Steering Law as they have the same width and length.
Therefore, it is expected that all four paths would take the same amount of time to steer through, but intuitively, we know that there would be some variation due to their differing shapes.

There have been many proposed extensions to expand the Steering Law for more complex tunnel shapes, such as widening/narrowing widths \cite{yamanaka_modeling_2016}, tunnels with corners \cite{pastel_measuring_2006}, curved paths \cite{zhai_human_2003, liu_effect_2010, nancel_modeling_2017, yamanaka_modeling_2019}, paths with sequential segments of varying widths \cite{liu_effect_2010, yamanaka_steering_2017, yamanaka_effectiveness_2022}, etc.
Specifically in the curved paths case, previous research established that curvature plays a significant role in steering tasks \cite{montazer_optimization_1988, zhai_human_2003}.
However, these studies are limited to scenarios involving constant curvature, such as circular paths or compound paths with circular arcs.
While these findings are valuable in demonstrating the effects of curvature, they may not accurately predict user performance in more complex scenarios where curvature can vary arbitrarily within a single path, such as 2D graphics segmentation or image labeling tasks.
The ability to predict the movement time and evaluate the \(ID\) of arbitrarily curved paths would also be helpful in determining level difficulties in video games such as Osu! \cite{osu}, Trombone Champ \cite{trombonechamp}, or even physical ones such as the wire and loop game.
% In our case, we are interested in applying the Steering Law in areas outside of user interface research.
Although they are typically evaluated on hand-based interactions, human performance models have valuable applications outside of user interface research in explaining the complexities of other motor control tasks such as vocal articular trajectories and tongue movements during speech production \cite{lammert_speed-accuracy_2018, kuberski_fitts_2021}.

Motivated by the need to enhance the predictive ability of steering models for a broader range of tasks such as navigation through arbitrarily curved paths, this study incorporates the overall curviness of a tunnel task as a critical factor.
It is important to note that while width is a traditional factor in movement models, our current study does not examine its effects and instead focuses instead on the implications of curvature.
Similar to how the width constraint is analogous to noise in Fitts' Law's physical interpretation of the task parameters, we hypothesize that the overall curviness of a path is also a source of noise, meaning that it would introduce variability in the overall movement time.
The contributions of this work are:
\begin{enumerate}
    \item We include a new factor, the total curvature parameter \(K\), into the Steering Law to account for varying curvatures, creating models that are more applicable to a broader range of steering tasks;
    \item We validated model variations and compared their fitness with models derived from previous work in a mouse-steering experiment with 20 participants on tunnel paths of varying lengths and curviness levels. We used a 3 \textbf{L} x 3 \textbf{K} x 15 repetitions x 20 participants design providing 2700 trials.
    Our results showed that the introduction of \(K\) significantly improves the Steering Law model movement time prediction;
\end{enumerate}

\section{Background}
In this section, we discuss the development and application of the Steering Law, focusing on its base formulation as well as the various extensions that have been proposed to address more complex paths.

\subsection{Fitts' and Steering Law Models}
Fitts' Law \cite{fitts_information_1954}, developed for reciprocal 1D target selection tasks, is given by:
\begin{align} 
    MT &= a + b \cdot ID \label{eq_fitts} \\ 
    ID &= \log_2\left(\frac{2L}{W}\right)
\end{align}
where \(a\) and \(b\) are empirically determined constants, \(L\) is the distance between the two targets and \(W\) is the width of the targets.
The model describes a linear relationship between \(ID\) and the movement time \(MT\).
It was later extended for 2D target tasks with different approach angles and target heights \cite{mackenzie_extending_1992, accot_refining_2003}.

By framing 2D steering tasks as a series of small pointing tasks along the length of a path, Accot and Zhai applied Fitts' Law to compute the cumulative effect of small target acquisition tasks and obtained the Steering Law \cite{accot_beyond_1997}.
The general form is expressed by the following equation for a path \(C\) of width \(W(s)\) at path length \(s\):
\begin{equation} \label{eq_steering_law_general}
    ID = \int_{C} \frac{ds}{W(s)}
\end{equation}
In cases where path width is constant, the \(ID\) can be rewritten as:
\begin{equation} \label{eq_steering_law_base}
    ID = \frac{L}{W}
\end{equation}
The \(ID\) of a steering task is also related to \(MT\) in the same linear relationship as Fitts' Law in Equation \ref{eq_fitts}.

\subsection{Steering Law Extensions}
In Accot and Zhai's original work, the Steering Law was validated for 3 different types of paths: a straight tunnel of constant width, a straight tunnel of decreasing width, and a spiral tunnel of increasing width \cite{accot_beyond_1997}.

In a later work, it was verified on circular paths of constant curvature \cite{accot_performance_1999}.
However, the data was analyzed separately from straight paths and the authors note that circular paths were significantly harder than straight paths, indicating that curviness may have a factor to play in the overall movement time.
% and for other applications such as locomotion tasks in virtual reality \cite{zhai_human_2003}.

Pastel's work extended the Steering Law to paths with corners, a form of path complexity involving a change in curvature that deviates from the traditional straight and uniform paths \cite{pastel_measuring_2006}.
His study modeled the task as a coupled motion of steering through a straight path and pointing at the corner, and found that the corner angle had a significant effect on the overall movement time.
However, since the study focuses on discrete corner angles, the applicability to paths with continuous curvature variations remains unexplored.

Liu et al. investigated the effects of path curvature and orientation in 3D interactive steering tasks through experiments where participants used a pen input device to navigate a virtual ball through paths of varying but fixed length, width, curvature and orientation \cite{liu_revisiting_2010}.
They found that the curvature and orientation of the circular arc path significantly influenced the overall steering time. This is in contradiction to the original Steering Law formulation, which only depends on the length and width parameters. Specifically for the curvature effect, they found that a model of the form:
\begin{equation} \label{eq_liu_og}
    \log_{10}(MT) = a + b \left[ \log_{10} \left( \frac{L}{W} \right) + c \ \kappa \right]
\end{equation}
where \(\kappa\) is the curvature of the circular arc provided the best fit to the experiment data.
\footnote{The logarithm used by Liu et al. was base 10. We keep the same form here for consistency.}

In another study, Liu and Liere extended their experiment to include scenarios where path properties change dynamically during a task by joining semicircular arcs of different widths and curvatures to form a complex tunnel shape \cite{liu_effect_2010}.
They compared the average steering velocities for each of the tunnel shapes, as well as for each type of tunnel section (J-joints and S-semicircles) within each tunnel shape.
Although they did not propose a model, they have found that the average velocity decreases when navigating through J sections and that curvature is highly correlated to the average velocity through S sections.
The sectioning and speed analysis approach they employed was useful in this case since the complex tunnel shape was composed of simple circular arcs.
This approach, however, may not extend well to scenarios where curvature is not uniform.

Drawing inspiration from the 2/3 Power Law relationship in curved trajectories, Nancel and Lank proposed an augmented model to accommodate paths of varying curvature and width \cite{nancel_modeling_2017} of the form:
\begin{equation} \label{eq_nl_og}
    MT = a + b' \int_{C} \frac{ds}{W(s) \cdot R(s)^{1/3}}
\end{equation}
where \(R(s)\) is path \(C\)'s radius of curvature at path length \(s\).
Similar to Liu and Liere's experiment, their steering task shape consisted of compound paths of joint circular arcs.
In their experiment however, each arc had smoothly varying widths, different arc angles and orientations.
It remains to be seen how the model performs for paths of arbitrary curvature.

Yamanaka and Miyashita investigated how the path curvature of simple circular arc paths affects movement time in pen-steering tasks \cite{yamanaka_modeling_2019}.
They fitted several regression models to predict the average speed and movement time based on width, length and the radius of curvature of the arc, and proposed a model of the form:
\begin{equation} \label{eq_ym_og}
    MT = a + b \frac{L}{W + c \ \frac{1}{R} + d \ \frac{W}{R}}
\end{equation}
Although the model showed good fit to the data, it was only validated on simple circular arcs.
% and the authors noted that it was only applicable to certain ranges of \(R\) and \(W\).

While previous extensions have expanded the Steering Law to accommodate diverse path types, they do not capture the steering behavior in paths with non-uniform, arbitrary curvatures, which our research aims to address directly.
Distinctively, we focus on fixed-width paths, allowing us to isolate the effects of curvature without the confounding influence of width changes.

\section{Model Derivation}
Building on the limitations identified in existing models, this section proposes a novel extension of the Steering Law that integrates the total curvature parameter \(K\) to account for the overall curviness of a path:
\begin{equation} \label{eq_total_curvature}
    K = \int_{C} \lvert \kappa(s) \rvert ds
\end{equation}

We return to the information theory basis of Fitts' Law and the signal and noise analogy and hypothesize that either a simple model of the form:
\begin{equation} \label{eq_add_K}
    MT = a + b \ L + c \ K
\end{equation} 
where \(K\) is an additive noise, or a case where the noise addition is logarithmic:
\begin{equation} \label{eq_add_logK}
    MT = a + b \ L + c \cdot \log_{2}(K+1)
\end{equation}
would improve on the model fitness of the original Steering Law.

For 2D parametric plane curves, the instantaneous curvature at point \(s\) along the curve is given by:
\begin{equation} \label{eq_instant_curvature}
    \kappa(s) = \frac{\lvert x'(s) \cdot y''(s) - y'(s) \cdot x''(s) \rvert}{(x'(s)^2 + y'(s)^2)^{3/2}}
\end{equation}

In a recent experiment, Kasahara et al. employed a sine-wave shape as one of the tunnel types in an experiment to refine throughput using effective parameters \cite{kasahara_better_2024}.
They found that the original Steering Law held up well with the sine wave shape and identified that the additional effect of curvature should be investigated for this shape.

In our experiment, we use sinusoidal waveforms created by adding simple sine-wave shapes of different frequencies and amplitudes as the steering task tunnel shape. 
Appendix \ref{appendix_A} presents the details of the curve construction and selection as well as a discussion on the curviness of sinusoidal paths.
Each curve is generated by:
\begin{align*}
    x(s) &= s \\
    y(s) &= \frac{a}{c} \sum_{i=0}^{c} sin(AM[i] \cdot \phi \cdot s)
\end{align*}

Together with Equations \ref{eq_total_curvature} and \ref{eq_instant_curvature}, we can compute the total curvature \(K\) of the paths.

In the case of straight paths, both Equations \ref{eq_add_K} and \ref{eq_add_logK} would be equivalent to the original Steering Law as the total curvature would be equal to 0.

It should be noted from Equations \ref{eq_add_K} and \ref{eq_add_logK} that we have omitted the \(W\) parameter.
A common behavior in steering through sharp corners identified in \cite{pastel_measuring_2006} is `corner cutting', which is the tendency of users to navigate wide corners by taking a shortcut, reducing the total movement time.
In our experiment, we use a small, constant width for all conditions to reduce corner cutting.
Further, to constrain the scope of our experiment, we are not examining the potential interaction of width and curvature, leaving that to future work. Instead, our study focuses on length and curvature.
The corner cutting behavior is discussed in the Limitations section as well.

In our analysis, we first compare the fitness of our model with the base Steering Law model by Accot and Zhai \cite{accot_beyond_1997} in Equation \ref{eq_steering_law_base}.
We also compare with the fitness of Nancel and Lank's model \cite{nancel_modeling_2017} in Equation \ref{eq_nl_og}, reformulated to remove the effect of width and using \(\kappa(s) = 1/R(s)\) as:
\begin{equation} \label{eq_nl}
    MT = a + b \ L \int_{C} \kappa(s)^{1/3} ds
\end{equation}
% Liu et al.\cite{liu_revisiting_2010} as well as
The model proposed by Yamanaka and Miyashita \cite{yamanaka_modeling_2019} was derived for circular arc paths, so we cannot directly compare it to our proposed models.
We will instead adjust their equation as if it was formulated for a path of arbitrary curvature by using \(\kappa(s) = \frac{1}{R}\) for a circular arc and plugging into Equation \ref{eq_total_curvature} to obtain \(K = \frac{L}{R}\). 
Rearranging Equation \ref{eq_ym_og} by absorbing the \(W\) parameter into the regression constants, we obtain:
\begin{equation} \label{eq_ym}
    MT = a + b \frac{L^2}{L + c \ K}
\end{equation}

Similarly for the model proposed by Liu et al. \cite{liu_revisiting_2010}, we modify Equation \ref{eq_liu_og} to absorb the width term into the regression intercept constant, use circular arc relationships to include the total curvature parameter,
and convert it to \(MT\) form from the original \(\log_{10}(MT)\) formulation:
\begin{equation} \label{eq_liu}
    MT = 10^{a + b \cdot \log_{10}(L) + c \frac{K}{L}}
    % \log(MT) = a + b \log(L) + c \frac{K}{L}
\end{equation}

All models derived in this section are summarized in Table \ref{table_params_mt} along with the model fitting results.

\section{Experiment}

\subsection{Setup and Data Acquisition}
\begin{figure}[h]
  \centering
  \includegraphics[width=\linewidth]{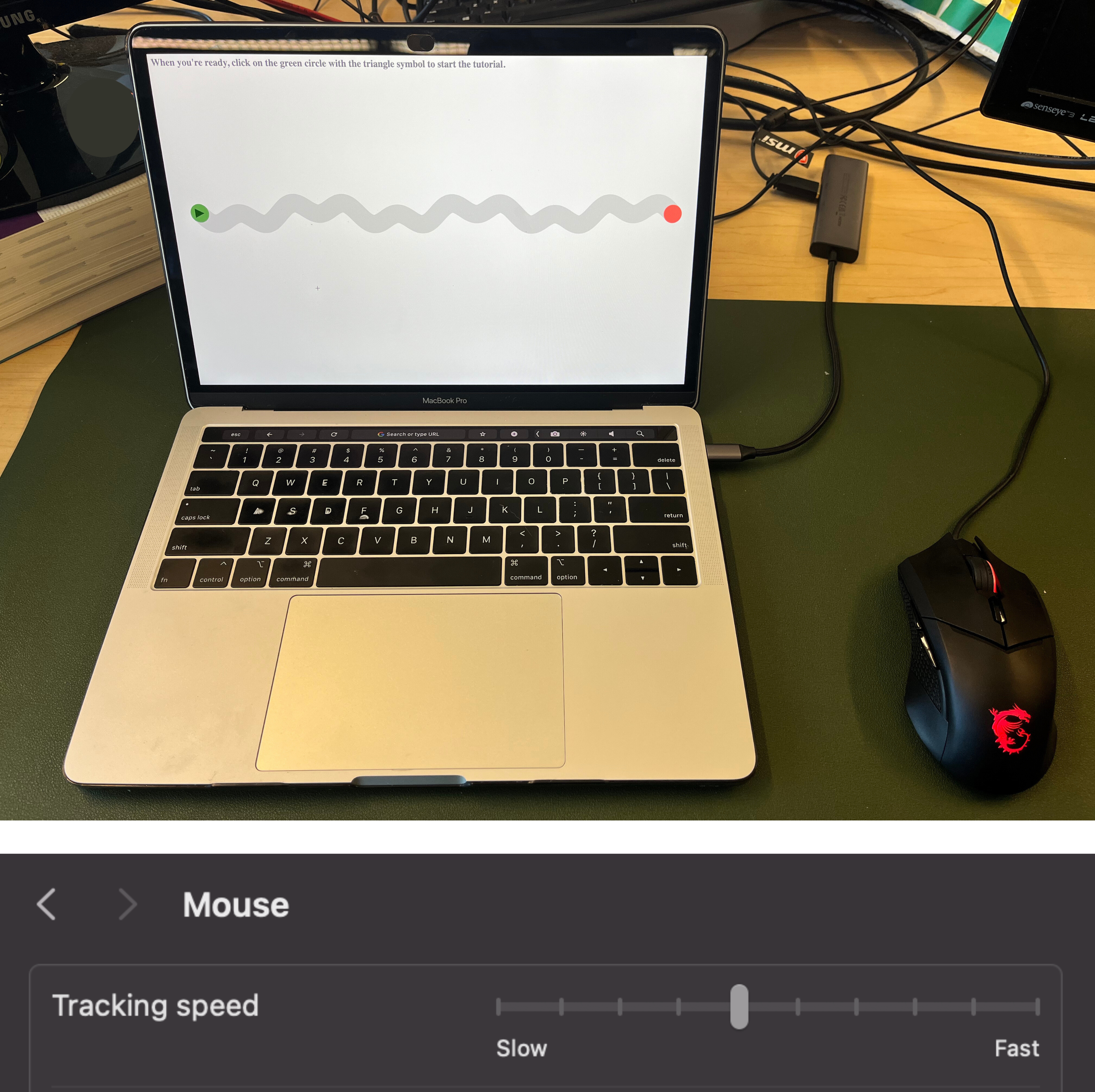}
  \caption{Hardware setup for the experiment and mouse sensitivity settings.}
  \label{fig_hw_setup}
  \Description{This figure is split into two parts. The top panel shows a photograph of the experimental hardware setup featuring a MacBook Pro with an open screen displaying the test interface. A wired gaming mouse is connected to the MacBook, positioned to its right side on a green desk mat. The bottom panel of the figure displays a screenshot of the mouse settings interface on the MacBook, detailing the adjustable tracking speed parameter which is set to the 5th notch from the left of the scrollbar.}
\end{figure}

The experiment was conducted on a Macbook Pro computer with the output displayed on a 13.3", 2560-by-1600 pixel display.
Participants used a 3200 DPI MSI Clutch GM08 mouse without additional weights as the sole input device to control the cursor.
Hardware acceleration was disabled for the mouse control and the sensitivity was fixed for all participants at the 5th notch (from the left of the scrollbar) in default Apple settings.
Mouse movements were polled and interpolated using cubic splines during post-processing to a rate of 200 Hz.
The setup is shown in Figure \ref{fig_hw_setup}.

For this study, the laptop display size was chosen for its common form factor and usage in everyday tasks, capturing natural user behavior.
Participants could adjust their seat height and viewing angle of the screen to be comfortable.
While outside the objective of this study, future work could extend our findings by controlling effective screen sizes and viewing angles to establish the movement time relationship between curviness, range of display sizes and viewing conditions as these are known to impact movement times \cite{kovacs2008perceptual}. Further, this type of study could determine whether large visual angles are associated with more efficient motor control strategy adoption when increased curviness makes the task more difficult, similar to \cite{kovacs2008perceptual}'s finding for Fitts' type tasks.

%We expect an influence since past research has shown that the size of the visual display influences the relationship between movement time and task difficulty in Fitts' tasks~\cite{kovacs2008perceptual}.
%% Specifically, a larger visual angle allows the human motor control system to employ more 
%% efficient control strategies in more difficult conditions \cite{kovacs2008perceptual}.

\subsection{Participants}
20 participants (9 women, 11 men) were recruited from the University of British Columbia's Vancouver campus for the experiment.
All participants had normal or corrected-to-normal vision.
4 participants were in the 19-24 age group and 16 participants were in the 25-34 age group.
In the background survey, 18 participants reported that they used the mouse daily, 1 participant used the mouse a few times a week, and 1 participant a few times a month.
All participants reported that they typically used the mouse using their right hand and subsequently performed the experiment using their right hand.
All participants successfully passed a tutorial that was performed at the start of the experiment to evaluate their speed and accuracy, as described below.
The experiment was conducted under approval from the university's behavioral research ethics board (Certificate Number: \textit{H24-01850}).

\subsection{Task}
\begin{figure*}[h]
  \centering
  \includegraphics[width=\linewidth]{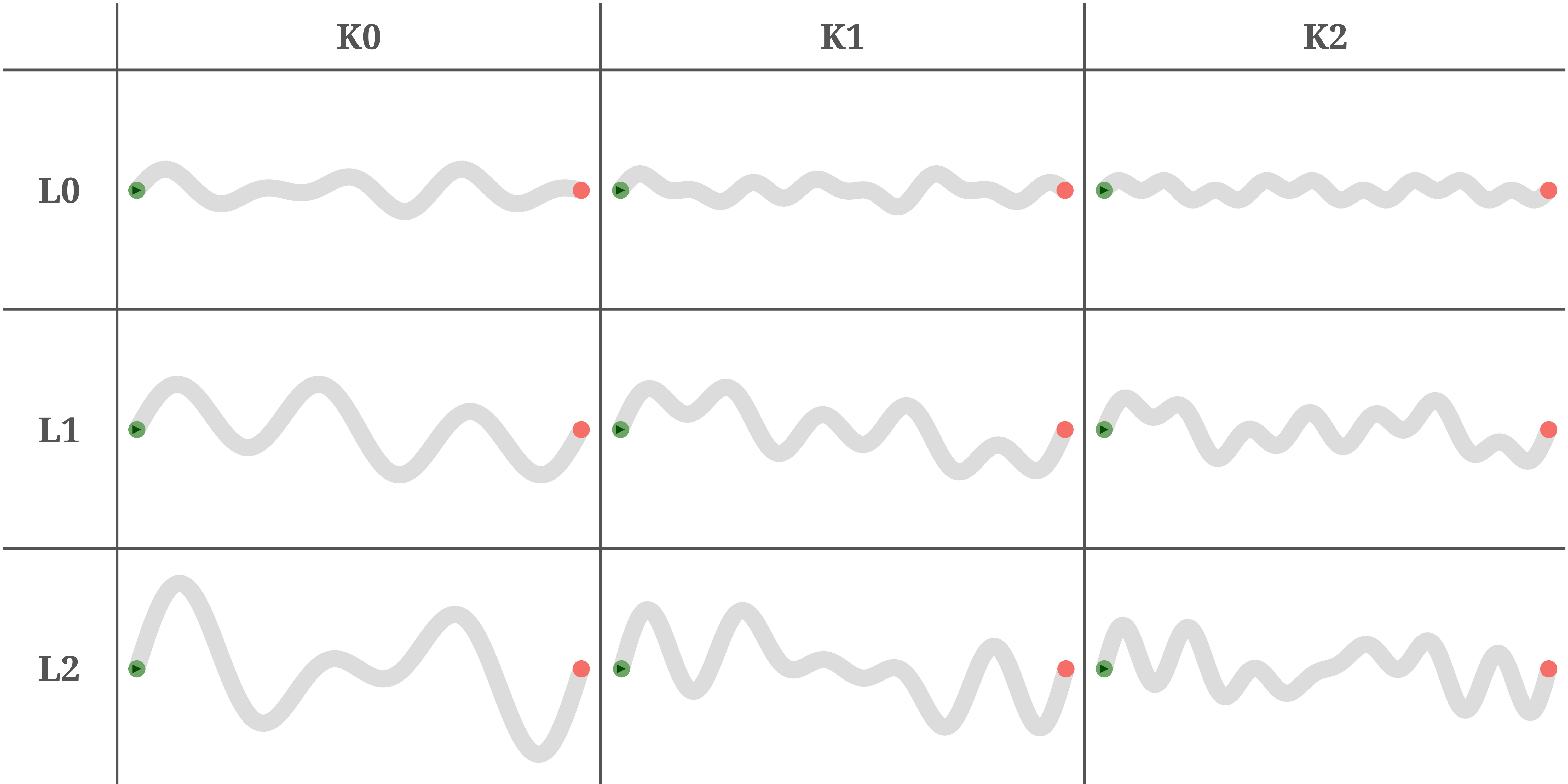}
  \caption{Tunnel tasks for each of the 9 trial types.}
  \label{fig_trial_example}
  \Description{This figure displays a grid of nine different tunnel paths arranged in three rows and three columns, representing all of the trials used to assess steering time. Each cell of the grid shows a gray path constructed from adding sine functions. The rows are labeled L0, L1, and L2, representing three levels of path length, and the columns are labeled K0, K1, and K2, representing three levels of curvature. Each path starts with a green circle and ends with a red circle, indicating the starting and ending points of the steering task.}
\end{figure*}

Participants were asked to move the provided mouse to steer a cursor as quickly and as accurately as possible through the displayed tunnels of varying conditions.
For each trial, the subject performed a series of actions: click the start button, steer the cursor through the tunnel to reach the other side, click the flag button, steer the cursor back through the tunnel to reach the starting point, then click the end button.
The action was reciprocal to account for potential directional effects and the steering direction was from left to right, then right to left.
The start and end buttons were always at the same location and spaced 1300 pixels apart.
A visual example of all of the trial types from the experiment is shown in Figure \ref{fig_trial_example}.

\subsection{Design and Procedure}
The study was divided into 2 phases: the tutorial phase and the experiment phase.
In the tutorial phase, participants were introduced to the same 9 types of paths that they would later encounter in the experiment phase.
The sequence of the tutorial tasks was pre-randomized and standardized across all participants.
Error circles were displayed on screen whenever the subject's cursor exited the tunnel's boundaries.
The participants' steering speed and accuracy were logged for a moving window of 8 trials.
The tutorial phase ended when a participant reached the practice threshold, or if they performed 30 trials without reaching this criterion.
In the latter case, they would not move onto the experiment phase.
The practice threshold was defined by maintaining a consistent cursor speed with a coefficient of variation below 0.15 and achieving an average error rate of less than 2 cursor exits per trial across the moving window.
These limits were set based on findings from prior pilot studies.

The experiment phase was divided into 27 blocks of 5 trials each, with a 15-second break between blocks.
The subject's cursor position and movement times were logged from the moment they clicked the start button until they completed the round-trip for each trial.
If a participant exited the defined path, the error was logged, but the participant was allowed to proceed without visual interruption or restarting.

A within-subjects study design with repeated measures was used.
Independent variables were total curvature and length, with 3 levels each. Tables \ref{table_lengths} and \ref{table_total_curvs} show a summary of the parameters for each trial.
For each combination of factors, subjects performed 15 repetitions which were divided into 3 blocks of 5 trials.
In total, each participant performed 135 trials and took approximately 1 hour for the entire experiment.

The sequence of blocks for the experiment was initially randomized and subsequently reversed for half of the participants to mitigate the learning effects throughout the experiment.
Complete counterbalancing was not feasible due to the size of our study.
Additionally, the orientation of the trials was pre-randomized.
A flipped trial involves mirroring the curve across the horizontal axis, preventing participants from anticipating the initial direction of the path at the start of each trial.
Figure \ref{fig_l0k0_flipped} in Appendix \ref{appendix_A} shows a visual example of `L0-K0' in its original and flipped orientations.

We fixed the width of all trials at 50 pixels to focus on the curvature effect.
It has been demonstrated in the past that corner cutting is a common behavior when navigating through corners, decreasing the overall movement time as the width of the path increases \cite{pastel_measuring_2006}.
In our case, the peaks of the sine wave are visually similar to corners.
Therefore, we chose a fixed width which limited this behavior but was not too small as to induce too many steering errors.

\begin{table}[h]    
    \renewcommand{\arraystretch}{1.2}
    \centering
    \caption{Lengths distribution per \textbf{L}}
    \begin{tabular}{|c|cc|}
        \hline
        \textbf{L} & Mean (px) & Std. (px) \\
        \hline
        0 & 1500.10 & 2.25 \\
        1 & 1882.33 & 2.23 \\
        2 & 2319.75 & 16.18 \\
        \hline
    \end{tabular}
    \label{table_lengths}

    \vspace{0.5cm}
    
    \renewcommand{\arraystretch}{1.2}
    \centering
    \caption{Total curvature distribution per \textbf{K}}
    \begin{tabular}{|c|cc|}
        \hline
        \textbf{K} & Mean (px) & Std. (px) \\
        \hline
        0 & 10.00 & 0.00 \\
        1 & 16.00 & 0.00 \\
        2 & 22.00 & 0.00 \\
        \hline
    \end{tabular}
    \label{table_total_curvs}
\end{table}

% \begin{table}[h]
    
% \end{table}

% *explain why std is 0 for curvature and nonzero for lengths
It should be noted that the lengths of the curves within each \textbf{L} level vary slightly due to the methods used in their construction.
In the following sections, \textbf{L} and \textbf{K} are used for ANOVA, while the actual values of \(L\) and \(K\) are employed in the linear regression model fitting process.

\section{Results and Discussion}

\subsection{Movement Time (\(MT\))}
We first perform repeated measures ANOVA with Greenhouse-Geisser correction for the movement time.
% The results are shown in Table \ref{table_anova_mt}.
Figure \ref{fig_results_all} a) illustrates the movement time for each trial averaged across \(n=15\) repetitions.
From the ANOVA results, we observed statistically significant main effects of \textbf{L} (\(F=216.55\); \(p < 0.001; \eta_{p}^{2}=0.402\)) and \textbf{K} (\(F=29.13\); \(p < 0.001; \eta_{p}^{2}=0.049\)).
We also found statistically significant but small interaction effects for \textbf{L x K} (\(F=3.95\); \(p < 0.05; \eta_{p}^{2}=0.008\)), which leads us to the addition of an extra interaction factor \(L \cdot K\) to Equations \ref{eq_add_K} and \ref{eq_add_logK}, creating the new models:
\begin{equation} \label{eq_compK}
    MT = a + b \ L + c \ K + d \ L \cdot K
\end{equation}
\begin{equation} \label{eq_compLogK}
    MT = a + b \ L + c \cdot \log_{2}(K+1) + d \ L \cdot K
\end{equation}

It is unclear what the information theoretic analogy for this interaction term would be. 
However, in a previous work by Yamanaka and Miyashita, they found that the inclusion of a significant interaction factor between the width and the reciprocal of the radius of curvature of a circular path resulted in higher model fitness than alternative models without it \cite{yamanaka_modeling_2019}.
Thus, we also investigate the effect of adding this significant interaction.

Table \ref{table_stats} shows a summary of the measurements for each of the trial types.
As the curviness of a path increases, the movement time also tends to increase.
The only exception is in trial `L2-K2', which has an unusual section as we discuss in the Limitations section.
This indicates that the steering task becomes increasingly more difficult the more curvy the task trajectory is, similar to findings from previous works \cite{montazer_optimization_1988, zhai_human_2003, liu_revisiting_2010, yamanaka_modeling_2019}.

% OLD: With tabularx
% \begin{table}[t]
%     \renewcommand{\arraystretch}{1.2}
%     \centering
%     \begin{tabularx}{\linewidth}{|@{}c *6{|>{\centering\arraybackslash}X}@{}|}
%     \hline
%     \multicolumn{1}{|c|}{} & \multicolumn{2}{c|}{Movement Time (ms)} & \multicolumn{2}{c|}{Out of Path Movement (\%)} & \multicolumn{2}{c|}{Average Speed (px/ms)}\\
%     \cline{2-7}
%     \multicolumn{1}{|c|}{Trial ID} & Mean & Std. & Mean & Std. & Mean & Std.\\
%     \hline
%     L0-K0 & 11932.85 & 2724.46 & 0.0021 & 0.0024 & 0.261 & 0.058 \\
%     L0-K1 & 13084.73 & 2439.85 & 0.0013 & 0.0020 & 0.232 & 0.041 \\
%     L0-K2 & 14809.09 & 3480.69 & 0.0005 & 0.0007 & 0.205 & 0.049 \\
%     L1-K0 & 15867.06 & 3521.73 & 0.0028 & 0.0031 & 0.250 & 0.052 \\
%     L1-K1 & 17152.55 & 4598.70 & 0.0027 & 0.0020 & 0.235 & 0.063 \\
%     L1-K2 & 18038.69 & 3587.48 & 0.0019 & 0.0021 & 0.215 & 0.041 \\
%     L2-K0 & 20092.95 & 4214.74 & 0.0052 & 0.0042 & 0.242 & 0.043 \\
%     L2-K1 & 21704.08 & 5663.71 & 0.0034 & 0.0029 & 0.229 & 0.057 \\
%     L2-K2 & 21432.13 & 4601.69 & 0.0018 & 0.0017 & 0.224 & 0.046 \\
%     \hline
%     \end{tabularx}
%     \caption{Summary of data distributions for the Movement Time (\(MT\)), Out of Path Movement (\(OPM\)) and Average Speed (\(v_{avg}\)) for each of the 9 trial types.}
%     \label{table_stats}
% \end{table}

\begin{table*}[h]
    \renewcommand{\arraystretch}{1.2}
    \centering
    \caption{Summary of data distributions for the Movement Time (\(MT\)), Out of Path Movement (\(OPM\)) and Average Speed (\(v_{avg}\)) for each of the 9 trial types.}
    \begin{tabular}{|c|p{2cm}|p{2cm}|p{2cm}|p{2cm}|p{2cm}|p{2cm}|}
    \hline
    & \multicolumn{2}{c|}{Movement Time (ms)} & \multicolumn{2}{c|}{Out of Path Movement (\%)} & \multicolumn{2}{c|}{Average Speed (px/ms)} \\
    \cline{2-7}
    Trial ID & \makebox[2cm][c]{Mean} & \makebox[2cm][c]{Std.} & \makebox[2cm][c]{Mean} & \makebox[2cm][c]{Std.} & \makebox[2cm][c]{Mean} & \makebox[2cm][c]{Std.} \\
    \hline
    L0-K0 & \makebox[2cm][c]{11932.85} & \makebox[2cm][c]{2724.46} & \makebox[2cm][c]{0.0021} & \makebox[2cm][c]{0.0024} & \makebox[2cm][c]{0.261} & \makebox[2cm][c]{0.058} \\
    L0-K1 & \makebox[2cm][c]{13084.73} & \makebox[2cm][c]{2439.85} & \makebox[2cm][c]{0.0013} & \makebox[2cm][c]{0.0020} & \makebox[2cm][c]{0.232} & \makebox[2cm][c]{0.041} \\
    L0-K2 & \makebox[2cm][c]{14809.09} & \makebox[2cm][c]{3480.69} & \makebox[2cm][c]{0.0005} & \makebox[2cm][c]{0.0007} & \makebox[2cm][c]{0.205} & \makebox[2cm][c]{0.049} \\
    L1-K0 & \makebox[2cm][c]{15867.06} & \makebox[2cm][c]{3521.73} & \makebox[2cm][c]{0.0028} & \makebox[2cm][c]{0.0031} & \makebox[2cm][c]{0.250} & \makebox[2cm][c]{0.052} \\
    L1-K1 & \makebox[2cm][c]{17152.55} & \makebox[2cm][c]{4598.70} & \makebox[2cm][c]{0.0027} & \makebox[2cm][c]{0.0020} & \makebox[2cm][c]{0.235} & \makebox[2cm][c]{0.063} \\
    L1-K2 & \makebox[2cm][c]{18038.69} & \makebox[2cm][c]{3587.48} & \makebox[2cm][c]{0.0019} & \makebox[2cm][c]{0.0021} & \makebox[2cm][c]{0.215} & \makebox[2cm][c]{0.041} \\
    L2-K0 & \makebox[2cm][c]{20092.95} & \makebox[2cm][c]{4214.74} & \makebox[2cm][c]{0.0052} & \makebox[2cm][c]{0.0042} & \makebox[2cm][c]{0.242} & \makebox[2cm][c]{0.043} \\
    L2-K1 & \makebox[2cm][c]{21704.08} & \makebox[2cm][c]{5663.71} & \makebox[2cm][c]{0.0034} & \makebox[2cm][c]{0.0029} & \makebox[2cm][c]{0.229} & \makebox[2cm][c]{0.057} \\
    L2-K2 & \makebox[2cm][c]{21432.13} & \makebox[2cm][c]{4601.69} & \makebox[2cm][c]{0.0018} & \makebox[2cm][c]{0.0017} & \makebox[2cm][c]{0.224} & \makebox[2cm][c]{0.046} \\
    \hline
    \end{tabular}
    \label{table_stats}
\end{table*}

\subsection{Out of Path Movement (\(OPM\))}
To examine the effects of the path parameters on the error rate, we use the \(OPM\), a measure introduced by Kulikov et al. \cite{kulikov_measuring_2005}.
This is calculated as the percentage of sampled points during a participant's trial that fall outside the boundaries of the drawn tunnel.
We found statistically significant main effects of \textbf{L} (\(F=23.062\); \(p < 0.001; \eta_{p}^{2}=0.112\)) and \textbf{K} (\(F=12.639\); \(p < 0.001; \eta_{p}^{2}=0.096\)).
We did not find statistically significant interaction effects for \textbf{L x K}.
Figure \ref{fig_results_all} b) shows the boxplot of the results, overlaid with the individual participant data.

\subsection{Average Speed (\(V_{avg}\))}
\begin{figure*}[h]
  \centering
  \includegraphics[width=\linewidth]{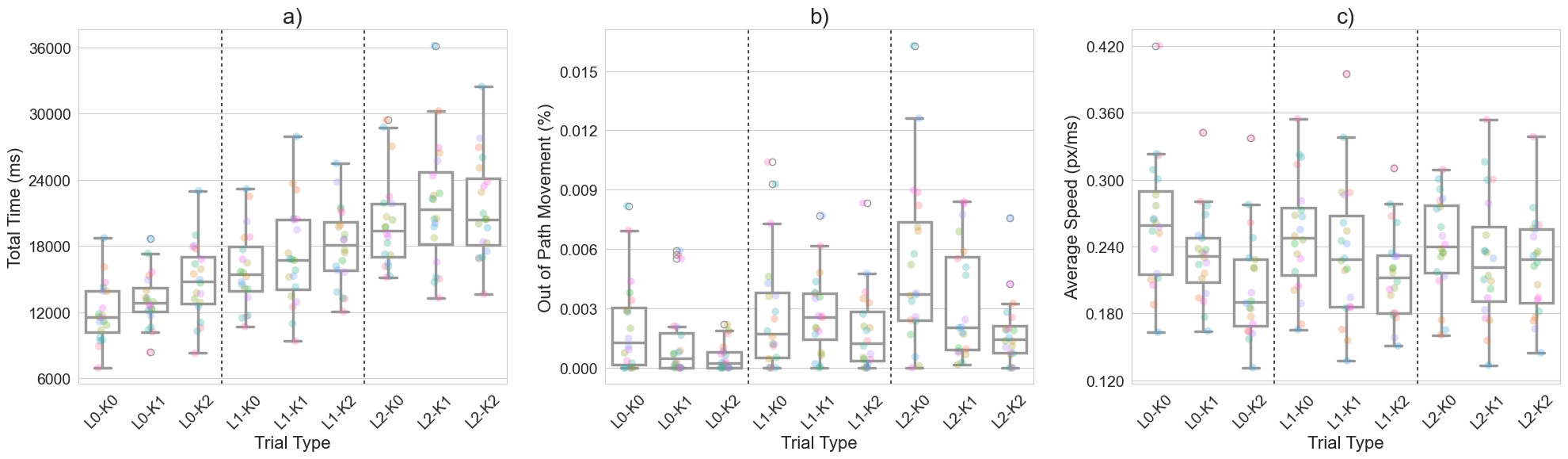}
  \caption{Plots of the a) Mean total time, b) Out of Path Movement, and c) Average Speed for each of the 9 trial types. Note that the boxes present the Interquartile Range (IQR) and median, while whiskers extend to 1.5 * IQR. Each color in the strip plot represents a different participant.}
  \label{fig_results_all}
  \Description{This figure displays three separate box plots illustrating different statistical measures across nine trial types of the experiment, highlighting trends and their alignment with ANOVA findings discussed in the Results and Discussion section. Subplot a) shows the Movement Time in milliseconds required to complete each trial type, ranging from around 6000 to 36000 ms. An upward trend in Movement Time is observed as K and L increase, indicating longer times required for more difficult tasks. Subplot b) shows the Out of Path Movement in percentages, ranging from 0\% to around 0.015\%. There is an upward trend as L increases, and a downward trend as K decreases, suggesting that participants tend to make more errors as the trials get longer, and also that curvier paths may encourage more precise steering. Subplot c) shows the average speed in pixels per millisecond, ranging from about 0.120 to 0.420 px/ms. There is a downward trend as K increases, reflecting slower navigation through more curved paths, but no significant trends across each of the levels of L. Each box plot represents the Interquartile Range with the median marked, and whiskers extending to 1.5 times the Interquartile Range. Colored dots distributed on top of each box plot in the form of a strip plot represent individual data points for different participants, with each participant assigned to a different color.}
\end{figure*}

The average speed is calculated as the amount of distance traveled by the cursor divided by the total time taken to complete the trial (in px/ms).
The results are shown in Figure \ref{fig_results_all} c).
We did not find statistically significant main effects for \textbf{L}, but found a large, significant effect for \textbf{K} (\(F=74.074\); \(p < 0.001; \eta_{p}^{2}=0.083\)).
We also found a small but statistically significant interaction effect for \textbf{L x K} (\(F=7.299\); \(p < 0.01; \eta_{p}^{2}=0.017\)).
We can note that as \(K\) increases, the average speed decreases.
This is in accordance with past studies \cite{montazer_optimization_1988, yamanaka_modeling_2019} and also the intuition that as a path gets curvier and more difficult, users slow down.

\subsection{Learning Effects}
Consistent with expectations, post-hoc analyses revealed a small but significant learning trend within each block (\(F=2.988\); \(p < 0.05; \eta_{p}^{2}=0.002\)) and a large, persistent learning effect over the course of the experiment (\(F=11.285\); \(p < 0.001; \eta_{p}^{2}=0.209\)).

To further investigate the learning process, we applied regression analysis for each trial type and participant to estimate the coefficients of the power law of practice, defined as:
\(MT(n) = a \cdot (n+1)^{-b}\), where \(n\) is the number of trials completed, \(a\) represents the initial performance level and \(b\) reflects the learning rate specific for each trial type and participant \cite{newell1981powerlaw}.
Our results indicated no statistically significant difference (\(p > 0.05\)) between the learning rates (\(b\)) across the trial types, indicating that all trial types were learned at comparable rates.

\subsection{Model Fitting}
\begin{figure*}[htb]
    \centering
    \vspace{0.2cm}
    \includegraphics[width=\textwidth]{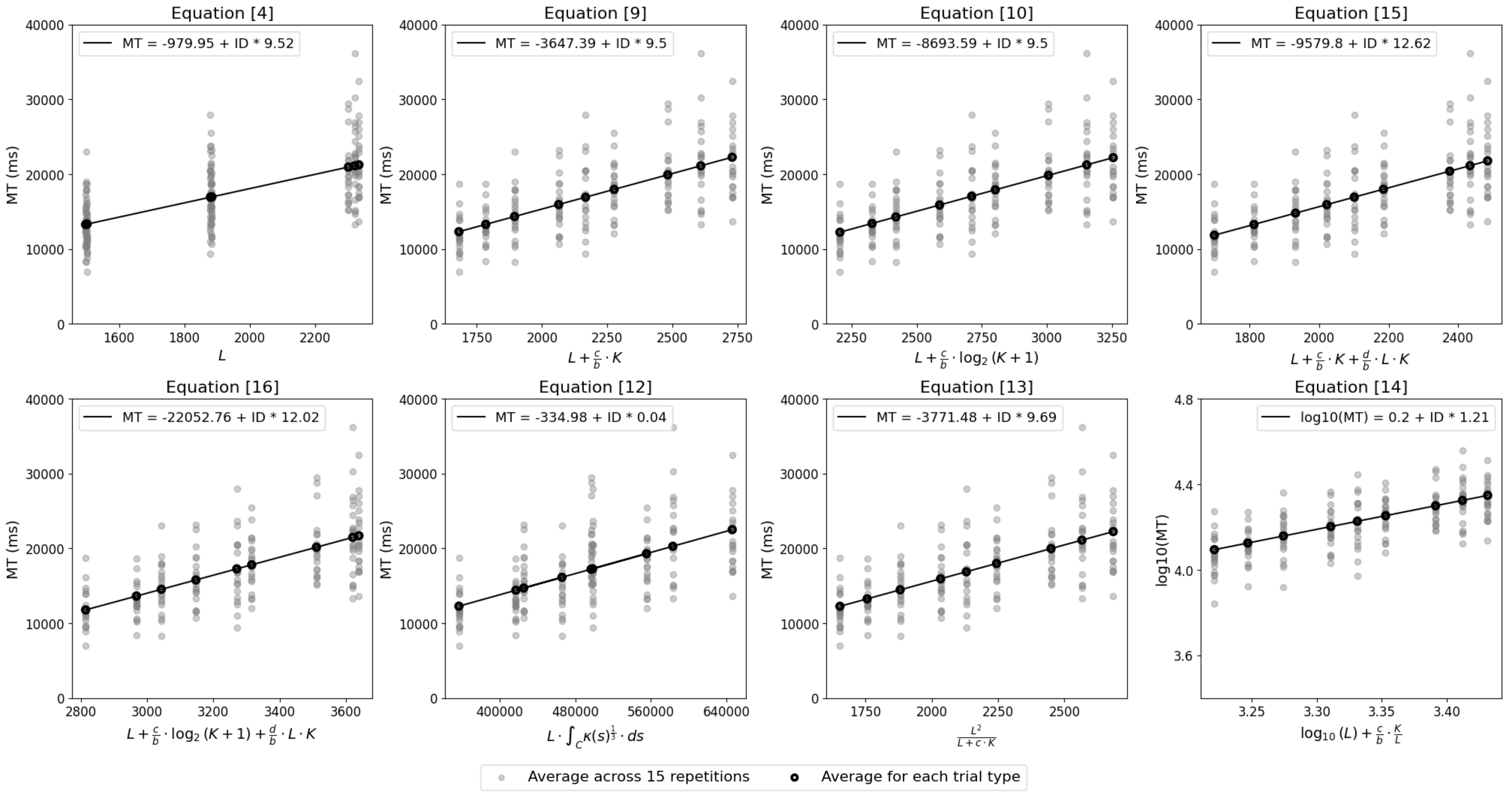}
    \caption{Visualization of model fitness of \(MT\) (or \(\log_{10} (MT)\) for Equation \ref{eq_liu}) plotted against \(ID\) for each model.}
    \label{fig_model_fitness}
    \Description{This figure shows eight scatterplots, each corresponding to a different regression model used to predict movement time (or log(MT) for the last subplot) based on the index of difficulty across various steering tasks. In each of the subplots, grey markers represent individual data points (average values of movement times across 15 repetitions for each trial type and participant), black markers represent the average movement time for each trial type, and the fitted line for each model. The scatterplots provide a visual comparison of model performances. Notably, plots corresponding to models lacking the total curvature parameter K exhibit more dispersed data points relative to their fitted lines, indicating a poorer fit. The visual dispersion illustrates the enhanced predictive accuracy of models that incorporate K, as these models show more data points clustering more closely to the fitted lines, indicating better alignment with the recorded data.}
\end{figure*}

% - w/ Path Parameters
Table \ref{table_params_mt} describes the linear regression coefficients, adjusted \(r^2\) values, and their Akaike Information Criterion (\(AIC\)) values for the \(MT\) models \cite{akaike_new_1974}.
Similar to other recent studies on Steering Law extensions comparing models \cite{yamanaka_steering_2019, yamanaka_modeling_2019, kasahara_better_2024}, we use \(AIC\) as the comparison metric as it balances the model fit and complexity.
A lower \(AIC\) value indicates a better model.
To check the fitness visually, Figure \ref{fig_model_fitness} shows the fitted models for \(MT = a + b \cdot ID\) on the average values for each trial type (\(N_{fit} = 9\)).
As well, the data points for each trial type and participant, averaged across 15 repetitions, are plotted in the same graph (\(N_{total} = 180\)).

\begin{table*}[h]
    \renewcommand{\arraystretch}{1.2}
    \centering
    \caption{Regression coefficients (with 95\% confidence intervals), adjusted \(r^2\) and \(AIC\) values for models predicting \(MT\). Note that *, **, and *** indicate \(p<0.05\), \(p<0.001\), and \(p<0.0001\) respectively.}
    \begin{tabular}{|c|p{4.5cm}||cccc||c|c|}
        \hline
        % \rowcolor{Graycolor}
        Eqn. & Model (\(MT\)) & \(a\) & \(b\) & \(c\) & \(d\) & Adj. \(r^2\) & \(AIC\)\\
        \hline
        \ref{eq_compLogK} & \(a + b \ L + c \cdot \log_{2}(K+1) + \ d \ L \cdot K\) & -22052.8*            & 12.0***      & 5238.3*            & -0.2* & 0.99 & 133.4 \\
                          &                                        & [-35672.9, -8432.7] & [9.4, 14.7] & [1890.6, 8586.0] & [-0.3, -0.0] &&\\[4px]
        \ref{eq_compK} & \(a + b \ L + c \ K + d \ L \cdot K\) & -9579.8*              & 12.6**       & 537.2*          & -0.2* & 0.99 & 134.1 \\ 
                       &                                       & [-15679.0, -3480.6] & [9.5, 15.8] & [175.6, 898.8] & [-0.4, -0.0] &&\\[4px]
        \ref{eq_ym} & \(a + b \ \frac{L^2}{L + c \ K}\)  & -3771.5*            & 9.7***       & -13.8** & & 0.981 & 139.0 \\ 
                    &                                    & [-6375.2, -1167.7] & [8.5, 10.9] & [-18.7, -8.9] & &&\\[4px]
        \ref{eq_add_logK} & \(a + b \ L + c \cdot \log_{2}(K+1)\) & -8693.6*             & 9.5***        & 1930.3* & & 0.981 & 139.0 \\ 
                          &                                       & [-12932.9, -4454.3] & [8.3, 10.7] & [1032.6, 2828.1] & &&\\[4px]
        \ref{eq_add_K} & \(a + b \ L + c \ K\) & -3647.4*           & 9.5***        & 170.1* & & 0.979 & 140.0 \\ 
                       &                       & [-6368.9, -925.9] & [8.3, 10.7] & [85.8, 254.3] & &&\\[4px]
        \ref{eq_liu} & \(10^{a + b \cdot \log_{10}(L) + c \frac{K}{L}}\)  & 0.2           & 1.2***      & 8.1* & & 0.978 & 140.7 \\ 
                     &                                                      & [-0.4, 0.8] & [1.0, 1.4] & [3.9, 12.3] & &&\\[4px]
        \ref{eq_steering_law_base} & \(a + b \ L\) & -980.0             & 9.5*** & & & 0.911 & 152.6 \\ 
                                   &               & [-5768.9, 3809.0] & [7.0, 12.0] & & &&\\[4px]
        \ref{eq_nl} & \(a + b \ L \int_{C} \kappa(s)^{1/3} ds\)  & -335.0             & 0.04** & & & 0.793 & 160.1 \\ 
                    &                                            & [-7777.8, 7107.8] & [0.0, 0.1] & & &&\\[3px]
        \hline
    \end{tabular}
    \label{table_params_mt}
\end{table*}

From the linear regression results in Table \ref{table_params_mt}, we can see that the models described by Equations \ref{eq_compK} and \ref{eq_compLogK} show the best fits in terms of \(AIC\) and adjusted \(r^2\).
While these two models contain the most free parameters, raising potential concerns about overfitting, \(AIC\) and adjusted \(r^2\) are criteria designed to penalize complexity.
To address any residual concerns, we perform cross-validation in the next section.
It should be noted that the regression coefficient for the interaction term is statistically significant (\(p < 0.05\)), which is in line with our ANOVA results.
This suggests that there is a need to include the \(K \cdot L\) term.

Models described by Equations \ref{eq_add_K}, \ref{eq_add_logK}, \ref{eq_ym} and \ref{eq_liu}, also exhibit strong performance in terms of \(AIC\) and adjusted \(r^2\).
All four models have \(AIC\) values within 2 units of each other, which indicates that they show comparable fitness.
Since they are within 10 units of \(AIC_{minimum}\) (133.4), we consider them to be valid representations of the data.
Most importantly, all four models incorporate the total curvature term \(K\) and outperform the base Steering Law model (Equation \ref{eq_steering_law_base}), {\it N.B.}, as it does not account for curvature variations in any way.
They also outperform the model described by Equation \ref{eq_nl}, which uses an alternative parameter based on the 2/3 Power Law for unrestrained movement \cite{nancel_modeling_2017, viviani_minimum-jerk_1995}.
In the original work by Nancel and Lank, this model was evaluated on compound paths with discontinuities in the width and curvature constraints not present in our paths, which may require additional adjustments in user behavior to transition between constraints.
In addition, the original Power Law relationship between the radius of curvature and movement speed is applicable to motion segments marked by landmarks such as cuspids and inflection points.
In Nancel and Lank's experiment paths, joints between circular arcs are inflection points, and their binning of radii of curvature adheres to the segmented motion assumption.
% Lastly, the power law introduced by Viviani et al. was developed for unrestrained movement such as handwriting, which may not extend well to visually-guided motion \cite{viviani_trajectory_1982}.
These differences likely contributed to its reduced performance on our experimental curves.
Overall, these findings strongly suggest that the inclusion of the parameter \(K\) improves the predictive accuracy of Steering Law models.

% Additionally, our results extend the work of Yamanaka and Miyashita \cite{yamanaka_modeling_2019}, as well as Liu et al. \cite{liu_revisiting_2010}, by demonstrating that their models can be extended to 2D paths of arbitrary curvature and constant width, treating uniform circular paths as a special case within the broader range of varying curvatures.
The models adapted from Yamanaka and Miyashita \cite{yamanaka_modeling_2019} as well as Liu et al. \cite{liu_revisiting_2010} underperformed slightly compared to our proposed models.
Since they were originally designed for simple or compound uniform circular paths, these models required modifications to fit our specific experimental conditions involving 2D paths of fixed width and arbitrary curvature.
We suspect that the transformation of these models---by assuming constant curvature as a specific instance of arbitrary curvature---might have left out terms that could further explain the variation in data observed in our study.
Despite these differences, the relatively strong performance of these adapted models confirms their validity and utility in contexts similar to our experimental setup.
% Additionally, in their original work, Yamanaka and Miyashita use average speed as a proxy for instantaneous speed, which is reasonable for their experiments on single circular arcs without inflection points.
% However, this assumption does not hold for our arbitrarily curved paths.
% Additionally, the Liu-based model may have also faced similar challenges as the Nancel and Lank model due to the lack of curvature discontinuities in the experiment path in our design.
% Both models were initially developed based on the instantaneous speed form of the Steering Law before being converted to the movement time form.
% This conversion may have worked well for shorter segments, but less for longer, non-uniform paths.

Returning once again to our proposed models derived from the signal and noise analogy, our model fitting results suggest that the curviness of a path may contribute to a logarithmic noise addition, as indicated by the lower adjusted \(r^2\) in Equations \ref{eq_add_logK} and \ref{eq_compLogK} compared to their counterparts in Equations \ref{eq_add_K} and \ref{eq_compK} 
The regression coefficient for the curviness term also has a smaller \(p\)-value in the logarithmic forms, suggesting a stronger effect.
However, the current study's range of \(K\) values may be too narrow to capture the full behavior of this relationship.
Expanding the range of \(K\) in future studies, especially the inclusion of \(K=0\) in straight paths, could provide further insights into the nonlinear noise addition.

By examining the regression coefficients for the models described by Equations \ref{eq_nl} and \ref{eq_liu}, we see that the \(a\) constants' 95\% confidence interval ranges include zero and non-significant \(p\)-values, which indicates that these models may perform better without this term.
By refitting the data to a modified version of Equation \ref{eq_nl} without an intercept (\(MT = b \ L \int_{C} \kappa(s)^{1/3} ds\)), we obtain \(b = 0.0347 [0.032, 0.037]\) with \(p<0.0001\), adjusted \(r^2\) value of 0.819 and an \(AIC\) value of 158.1, exhibiting an improvement in model fitness.
Then, doing the same with Equation \ref{eq_liu} (\(MT = 10^{b \log_{10}(L) + c \frac{K}{L}}\)), we obtain \(b = 1.267 [1.258, 1.276]\), \(c = 8.887 [5.521, 12.253]\) with \(p<0.0001\), adjusted \(r^2\) value of 0.978 and \(AIC\) value of 139.8.
The slight improvement may simply be due to the model using one less parameter.
The improved forms of these two models are used in the cross-validation.

\subsection{Cross-Validation of Models}
We conduct a form of 5-fold cross-validation to further assess the potential for overfitting, despite the ability of \(AIC\) and adjusted \(r^2\) measures to balance out model complexity with fitness.
During the model fitting process, our dataset comprised of 9 data points, one for each combination of \textbf{L} and \textbf{K}, with each point obtained from the average movement time across all participants and for 15 repetitions of each trial type.
For the cross-validation analysis, each of the models is fitted 5 times, each time using 9 datapoints computed from the average of 12 of the 15 repetitions.
For each fold, the models are tested on the average of the last 3 trials left out.
The average Root Mean Square Error (\(RMSE\)) values across the 5-fold validation are shown in Table \ref{table_rmse}.

The results indicate that the models incorporating the \(L \cdot K\) interaction term (Equations \ref{eq_compK} and \ref{eq_compLogK}) consistently have the highest performance, as indicated by the lowest \(RMSE\) values.
Based on the cross-validation results, we can conclude that overfitting is not a concern in these models, even with the inclusion of additional terms.

% OLD: Tabularx
% \begin{table}[h]
% \renewcommand{\arraystretch}{1.3}
% \caption{Average Root Mean Square Error values computed across 5-fold cross validation}
% \begin{tabularx}{250px}{|c|X||c|}
%     \hline
%     Eqn.                       & Model (\(MT\))                                          & \(RMSE\) (ms)\\
%     \hline
%     \ref{eq_compLogK}          & \(a + b \ L + c \cdot \log_{2}(K+1) + d \ L \cdot K\)  &  372.33\\
%     \ref{eq_compK}             & \(a + b \ L + c \ K + d \ L \cdot K\)                  &  381.18\\
%     \ref{eq_ym}                & \(a + b \ \frac{L^2}{L + c \ K}\)                      &  480.76\\
%     \ref{eq_add_logK}          & \(a + b \ L + c \cdot \log_{2}(K+1)\)                  &  480.88\\
%     \ref{eq_add_K}             & \(a + b \ L + c \ K\)                                  &  498.62\\
%     \ref{eq_liu} (without a)   & \(10^{b \cdot \log_{10}(L) + c \frac{K}{L}}\)          &  535.36\\
%     \ref{eq_steering_law_base} & \(a + b \ L\)                                          &  967.33\\
%     \ref{eq_nl} (without a)    & \(b\ L \cdot \int_{C} \kappa(s)^{1/3} ds\)             &  1435.43\\
%     \hline
% \end{tabularx}
% \label{table_rmse}
% \end{table}

\begin{table}[h]
\renewcommand{\arraystretch}{1.2}
\centering
\caption{Average Root Mean Square Error values computed across 5-fold cross validation}
\begin{tabular}{|c|p{4.5cm}||c|}
    \hline
    Eqn.                       & Model (\(MT\))                                         & \(RMSE\)\\
    \hline
    \ref{eq_compLogK}          & \(a + b \ L + c \cdot \log_{2}(K+1) + d \ L \cdot K\)  &  372.33\\[4px]
    \ref{eq_compK}             & \(a + b \ L + c \ K + d \ L \cdot K\)                  &  381.18\\[4px]
    \ref{eq_ym}                & \(a + b \ \frac{L^2}{L + c \ K}\)                      &  480.76\\[4px]
    \ref{eq_add_logK}          & \(a + b \ L + c \cdot \log_{2}(K+1)\)                  &  480.88\\[4px]
    \ref{eq_add_K}             & \(a + b \ L + c \ K\)                                  &  498.62\\[4px]
    \ref{eq_liu} (without a)   & \(10^{b \cdot \log_{10}(L) + c \frac{K}{L}}\)          &  535.36\\[4px]
    \ref{eq_steering_law_base} & \(a + b \ L\)                                          &  967.33\\[4px]
    \ref{eq_nl} (without a)    & \(b\ L \cdot \int_{C} \kappa(s)^{1/3} ds\)             &  1435.43\\[3px]
    \hline
\end{tabular}
\label{table_rmse}
\end{table}

\subsection{Limitations and Future Work}
\begin{figure}[h]
  \centering
  \includegraphics[width=\linewidth]{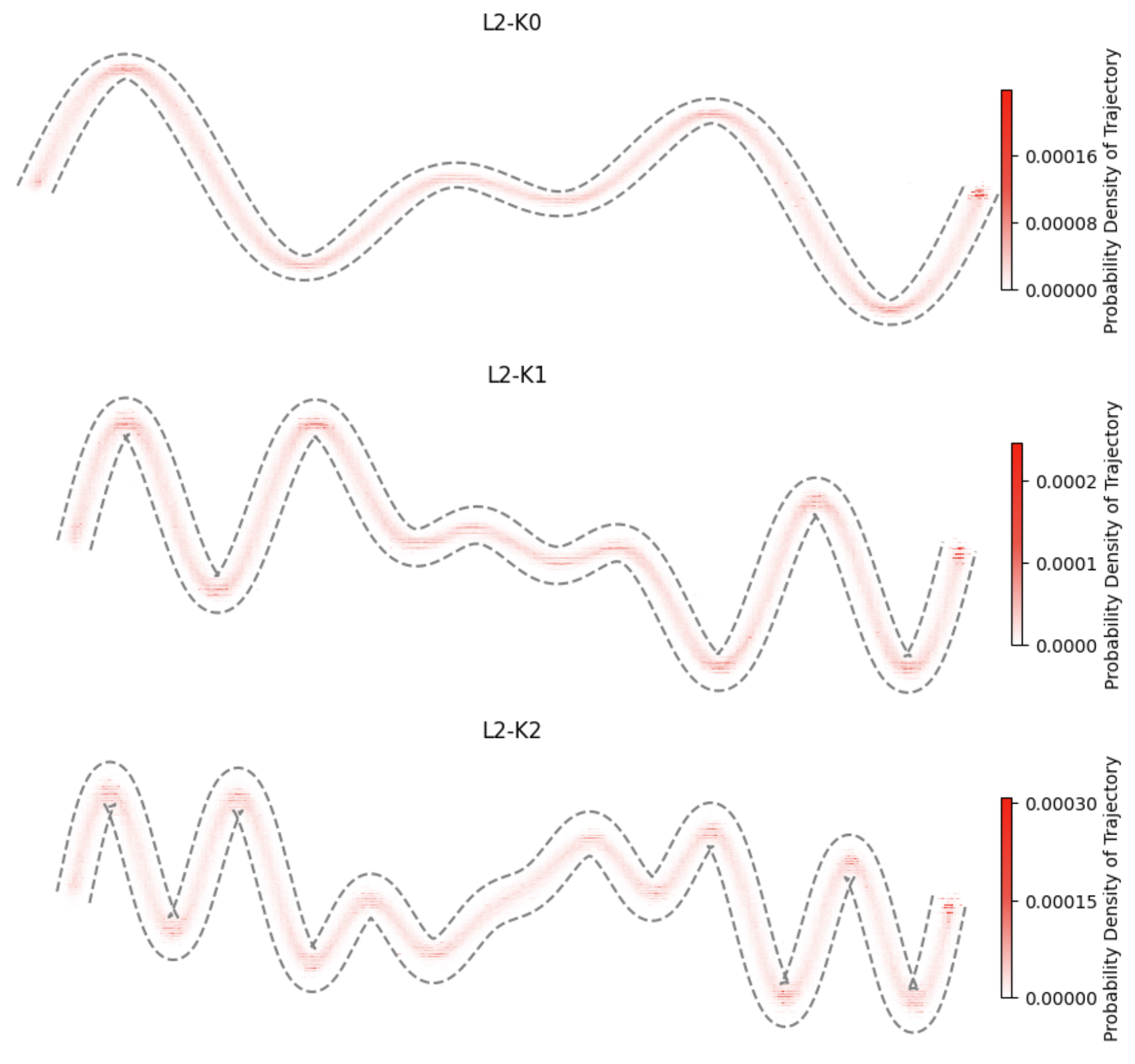}
  \caption{Heatmap of trajectory frequencies computed over 300 repetitions for each of the \textbf{L2} trial types.}
  \label{fig_heatmaps}
  \Description{This figure presents three heatmaps of trajectory frequencies for trials L2-K0, L2-K1, and L2-K2. Each heatmap shows the outline of the path and the saturation inside the path represents the frequency of trajectory paths taken by participants. Bright red areas indicate where participants spend more time. The frequencies, computed from 300 repetitions for each trial type, reveal distinct patterns: more saturated spots typically appear at the local peaks and less saturated spots at points of inflection within each curve. This pattern suggests that participants adjust their steering most significantly at these landmarks, either accelerating or decelerating. Notably, an exception occurs in the middle section of L2-K2, which is characterized by small undulations. As detailed in the Limitations and Future Work subsection, rather than slowing down at these local peaks, participants often shortcut these sections by making a straight line across them, demonstrating the corner cutting behavior. This observation highlights a particularity in the participants’ steering strategies that is not fully explained by the proposed model and K parameter.}
\end{figure}

The main limitation of our study is the exclusion of the width parameter as one of the independent variables in our trials.
We expect that the corner cutting behavior can be better understood through analyzing tunnels of different widths.
For instance, as seen in Figure \ref{fig_results_all} a), the trial `L2-K2' had a lower average movement time than `L2-K1' despite the increase in curviness.
We suspect that this behavior is due to the very small undulations in the middle portion of the `L2-K2' trial that allows the section to be bypassed as a straight tunnel task.
A visual inspection of the plotted heatmaps of the trajectory frequencies of all 300 repetitions of the \textbf{L}2 trials, shown in Figure \ref{fig_heatmaps}, supports this hypothesis.
The participant trajectories do not follow the undulations in the middle section of `L2-K2', suggesting that the large width tolerance of the tunnel allowed participants to `cut corners'.
We suspect that a narrower tunnel would result in longer movement times.

Additionally, the heatmaps highlight that areas of highest curvature (i.e. the sharpest corners) are more brightly colored than the rest.
This may correspond to the `corner aiming' behavior identified in \cite{pastel_measuring_2006}, which happens in narrow corners.
The corner cutting and aiming effect could be interpreted as participants using a perceived motion model, preemptively adjusting their movement to avoid unnecessary corrections, reducing the noise transmitted through the human channel (i.e. the curviness of the path) and thereby decreasing the \(ID\) of the overall task.
Interestingly, incorporation of predictive movement models and linguistic and physical constraints for reducing the \(ID\) of the complex trajectory movement tasks, like tongue movements in speech, may be a strategy for reducing motor control load while speaking \cite{gick_speaking_2017}.

In future work, incorporating {\it effective} parameters into the models such as \(W_e\) and \(L_e\), which account for actual user behavior, could improve the movement time prediction as demonstrated in previous studies \cite{kulikov_measuring_2005, zhou_speed-accuracy_2009, kasahara_better_2024}.
Specifically, \(W_e\) is computed from the standard deviation \(\sigma\) of movement coordinates perpendicular to the centerline of the tunnel as: \(W_e = 4.133 \ \sigma\).

\(W_e\) represents both the utilization rate of the entire width of the tunnel as well as a measure of error.
This is especially important in the information-theoretic analogy to understand how the variability in noise (width and curviness constraints) perturbs the movement \cite{mackenzie_fitts_1992}.
Understanding how \(W_e\) interacts with the curviness of the tunnel would provide more insight on corner cutting/aiming behaviors.

For a future study on the effects of curviness and width, an additional parameter that could be used is the curviness of the shortest path through the tunnel between the two endpoints.
\(K_{min}\) could potentially account for sections that can be bypassed as straight tunnels as well as the curviness of the shortest edge of sharp corners.

Another limitation of the study is the restricted range of path lengths evaluated.
Past research suggests that the Steering Law has operational constraints at different scales of motion \cite{accot_scale_2001}.
Future work would involve evaluating the model on both shorter and longer paths to better understand and decouple the effects of scaling.
% A future work identified previously is to determine whether the addition of \(K\) as a noise parameter in the information theory analogy is linear or logarithmic.
% The two models we have compared have very similar fitness.
% Perhaps a model with a width term would reveal this.

% - no effect of orientation (only left-right) and only 1 back and forth
% - length of curves are slightly different in each L level -> can lead to small deviations in the independence assumption
% - eqn. 13 showed really good r2, but not examined in depth. Shows that log(MT) might improve other models too

\section{Conclusion}
In this paper, we introduced the total curvature parameter \(K\) as a significant extension to the Steering Law to better account for the curviness of constant-width paths in movement time prediction and human behavior modeling.
Our study demonstrated that the inclusion of \(K\) improves model fitness, as indicated by the higher adjusted \(r^2\), as well as lower \(AIC\) and \(RMSE\) values when fitted and cross-validated on the data.
We suggest the following form:
\begin{align*}
    MT &= a + b \ L + c \cdot \log_{2}(K+1) + d \ L \cdot K 
\end{align*}
which includes a significant interaction term \(L \cdot K\) between path length and curviness.

We have also adapted models originally developed for uniform circular paths to be applicable on paths of fixed width and varying curvature.
% the models proposed by previous work on uniform circular paths and showed that fixed-width paths of varying curvature can be treated as a general case, with uniform circular paths as a specific instance.
By excluding width and incorporating the total curvature parameter \(K\), the models derived from prior work by Yamanaka and Miyashita (Equation \ref{eq_ym}) and Liu et al. (Equation \ref{eq_liu}) have demonstrated good fitness, effectively capturing a varied spectrum of curved path scenarios.
% when reformulated with the \(K\) parameter.

The extensions to the Steering Law and the inclusion of path curviness can help designers understand and optimize user paths in complex interfaces, such as in menu selections, accessibility design, lasso operations, image segmentation tasks, etc.
Beyond user interface design, it has applications in areas such as motor control, virtual navigation, game level difficulty evaluation, and extends into understanding complex motor tasks like speech production.
% Future work will focus on refining these models by incorporating other factors like path width and further exploring the interaction effects between length and curvature.
% These improvements would enhance the utility of the Steering Law for predicting movement time in a wide variety of real-world tasks.

\begin{acks}
We would like to thank the participants of the experiment for their valuable time and contributions.
We also thank our colleagues and friends for the insightful discussions.
\end{acks}

\bibliographystyle{ACM-Reference-Format}
\bibliography{se-paper}

\newpage
\appendix

\section{Experiment Curve Construction and Parameter Choice} \label{appendix_A}

As described in the Model Derivation section, each sinusoidal curve is generated by:
\begin{align*}
    x(s) &= s \\
    y(s) &= \frac{a}{c} \sum_{i=0}^{c} sin(AM[i] \cdot \phi \cdot s) \\
    \phi &= \frac{np \cdot 2 \pi}{x_{max}}
\end{align*}

where \(a\) is the amplitude, \(c\) is the number of sinusoidal functions used in the overall construction of the curve (ranging from 1 to 3), \(AM\) is a list of \(c\) angle multipliers, \(np\) is the number of periods, and \(x_{max}\) is the maximum value of \(x\) (fixed at 1300 px).
Table \ref{table_lengths_curvatures} summarizes the lengths and total curvatures of each of the trial types.

\begin{table}[h]
    \centering
    \renewcommand{\arraystretch}{1.2}
    \caption{Lengths and total curvatures of each of the 9 trial types.}
    \begin{tabular}{|c|c|c|}
        \hline
        Trial ID & Length (px) & Total Curvature \\
        \hline
        L0-K0 & 1502 & 10 \\
        L0-K1 & 1498 & 16 \\
        L0-K2 & 1500 & 22 \\
        L1-K0 & 1885 & 10 \\
        L1-K1 & 1880 & 16 \\
        L1-K2 & 1882 & 22 \\
        L2-K0 & 2303 & 10 \\
        L2-K1 & 2322 & 16 \\
        L2-K2 & 2335 & 22 \\
        \hline
    \end{tabular}
    \label{table_lengths_curvatures}
\end{table}

To identify possible candidates for each of the trial types, we conduct a grid search over the sinusoidal function parameters \(c\), \(AM\) and \(np\), and the targeted curviness \(K\).
For each combination of these parameters, we solve for the corresponding amplitude \(a\), compute the total length of the resulting curve, and record the parameter sets that produce curves in range with the desired lengths.
Finally, through visual inspection of the resulting curves and pilot trials, we select the most appropriate curve for each \(L\) and \(K\) combination, specifically choosing those that minimize corner cutting behavior and prioritizing matching the lengths of the curves within each designated category to ensure consistency across trials.

In the case of `L2-K2', alternative curves generated during the search process for this trial type include those with lengths of 2246, 2421 and 2441 pixels.
These were significantly outside the target range for the \textbf{L}2 category, with `L2-K0' and `L2-K1' having lengths of 2303 and 2322 pixels, respectively.
Given the importance of length consistency in our experimental design, curves that did not meet this criterion were excluded from consideration, regardless of their other characteristics.

\begin{figure}[!h]
  \centering
  \includegraphics[width=0.75\linewidth]{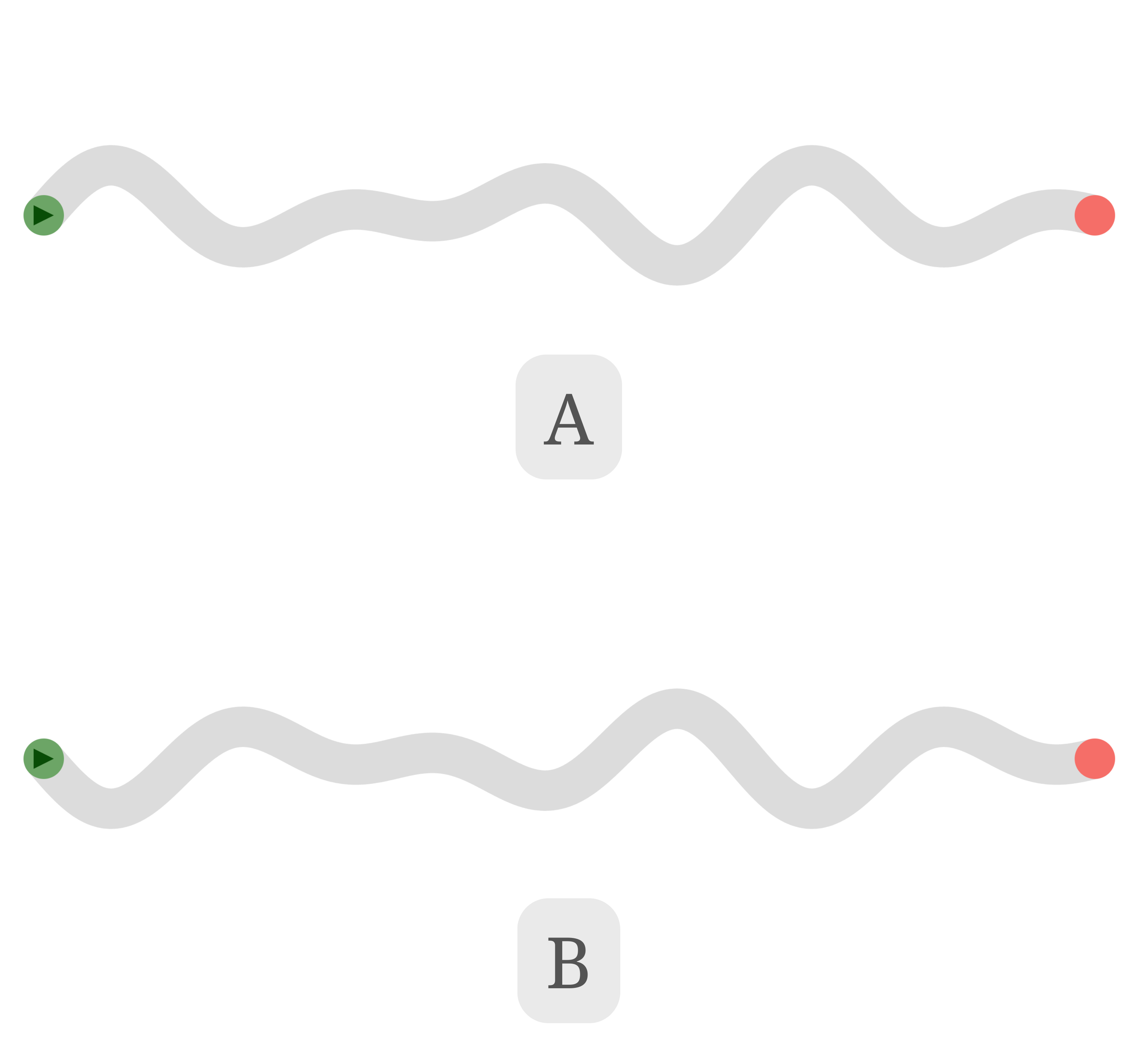}
  \caption{The two possible orientations of trial `L0-K0'. A) Original orientation; B) Flipped orientation.}
  \label{fig_l0k0_flipped}
  \Description{This figure displays the two possible orientations of trial L0-K0. Each path starts with a green circle and ends with a red circle, indicating the starting and ending points of the steering task. Panel A shows the original version with the curve initially ascending. Panel B shows the flipped version of the original curve across the horizontal axis with the curve initially descending.}
\end{figure}

In addition, another parameter that determines the visual presentation of the trial is its orientation.
Figure \ref{fig_l0k0_flipped} shows the two possible orientations of trial `L0-K0'.

A note on sinusoidal waveforms is that while one could attempt to use its energy parameter \(E\) as an alternative to \(K\) to represent its overall curviness.
The energy of a finite sine signal \(y(s) = A \cdot sin(\frac{2 \pi}{T} s)\) over 1 period \(T\) is \(E = \frac{A^2 \cdot T}{2}\).
As the amplitude \(A\) increases, the height of the curve described by \(y(s)\) also increases, which is in accordance with the idea that \(E\) becomes larger as the path becomes curvier.
However, as the frequency \(f = 1/T\) of the signal increases, say, by double, \(T\) is halved.
If we compute the new energy of the function from 0 to 2T, it would still have the same energy as before, despite the fact that the curve would be subjectively curvier due to the extra bends.
For this reason, we did not choose \(E\) to represent the curviness of a path.

\end{document}